\documentclass[12pt]{iopart}

\usepackage{graphicx}
\usepackage[usenames, dvipsnames]{color}
\usepackage{amssymb}
\usepackage{hyperref}
\begin{document}
\title[AWAKE readiness ...]{AWAKE readiness for the study of the seeded self-modulation of a 400\,GeV proton bunch}

\author{ The AWAKE collaboration\\
P.	Muggli$^{1,2}$,
 E.	Adli$^3$, 
 R.	Apsimon$^4$, 
 F. Asmus$^1$, R.	Baartman$^5$,
  A.-M. Bachmann$^{1,2,6}$, 
M.	Barros Marin$^2$, F.	Batsch$^{1,2,6}$, J.	Bauche$^2$, V. K.	Berglyd Olsen$^3$
% ...
}

\address{
$^1$Max Planck Institute for Physics, M{\"u}nchen, Germany\\
$^2$CERN, Geneva, Switzerland\\
$^3$University of Oslo, Oslo, Norway\\
$^4$Lancaster University, Lancaster, UK\\
$^5$TRIUMF, Vancouver, Canada\\
$^6$Technical University Munich, Munich, Germany\\
% ...
}

\ead{muggli@mpp.mpg.de}
\vspace{10pt}
\begin{indented}
\item[]July 2017
\end{indented}

\begin{abstract}
AWAKE is a proton-driven plasma wakefield acceleration experiment. %
We show that the experimental setup briefly described here is ready for systematic study of the seeded self-modulation of the 400\,GeV proton bunch in the 10\,m-long rubidium plasma with density adjustable from 1 to 10$\times10^{14}$\,cm$^{-3}$. %
We show that the short laser pulse used for ionization of the rubidium vapor propagates all the way along the column, suggesting full ionization of the vapor. %
We show that ionization occurs along the proton bunch, at the laser time and that the plasma that follows affects the proton bunch. %
\end{abstract}

%
%
%
%
%

\section{Introduction}

Plasma-based accelerators (PBAs) hold the promise of accelerating particles with a higher gradient than radio-frequency accelerators.  %
 A PBA can be driven by an intense laser pulse~\cite{bib:tajima} or a relativistic particle bunch~\cite{bib:chen}. %
A PBA acts as an energy transformer, extracting energy from the driver, transferring it to the the wakefields sustained by the plasma electrons, and then to the witness bunch. %
The energy carried by the drive bunch is the maximum energy that can be transferred to the witness bunch. %
Relativistic proton bunches produced by synchrotrons such as the SPS, LHC, Tevatron, RHIC, etc. carry many kilojoules of energy. %
Proton bunches are therefore interesting drivers to accelerate particles to very high energies in a single plasma. 

Plasma wakefields are fields sustained by an electron density perturbation in an otherwise neutral plasma. %
In one dimension, plasma wakefields are sustained by an electro-static relativistic electron or Langmuir wave. %
The driving of large amplitude wakefields requires a relativistic particle bunch (i.e., with velocity $v_b\cong c$), or a laser pulse, with transverse and longitudinal dimensions on the order of the cold plasma skin depth $c/\omega_{pe}$. %
This is usually expressed as $k_{pe}\sigma_r\le1$ and $k_{pe}\sigma_z\le1$. 
Here $k_{pe}=\omega_{pe}/c$ is the wavenumber and $\omega_{pe}=\sqrt{n_ee^2/\epsilon_0m_e}$ is the angular frequency of a relativistic plasma wave sustaining the wakefields in a plasma with electron density $n_e$. %
The bunch or pulse rms transverse and longitudinal sizes are $\sigma_r$ and $\sigma_z$, respectively. %
Focusing provides the small transverse size. %
Compression usually provides the small longitudinal size or length. %
However, compression leads to high bunch current or laser pulse intensity, thereby limiting the charge of the particle bunch or the energy of the laser pulse, that is limiting the energy they carry. %
Short particle bunches and laser pulses ($<$1\,ps or 300\,ps) usually carry an energy lower than 100\,J. %

The CERN accelerator complex provides relativistic proton ($p^+$) bunches that carry 20\,kJ (Super Proton Synchrotron, SPS, 3$\times 10^{11}p^+$/bunch, 400\,GeV/$p^+$, $\gamma_0\cong$427) to 112\,kJ (Large Hadron Collider, LHC, $10^{11}p^+$/bunch, 7\,TeV/$p^+$, $\gamma_0\cong$7500). %
However, these bunches are long, with $\sigma_z$=10-12\,cm. %
AWAKE aims to use these bunches to drive wakefields in plasmas. %
In the following we use the AWAKE baseline parameters given in Table~\ref{table:expparams}, unless otherwise specified.

The longitudinal electric field sustained by plasma wakefields can reach an amplitude on the order of the cold plasma wave-breaking field~\cite{bib:dawson}: $E_{WB}=m_ec\omega_{pe}/e\cong\sqrt{n_e[10^{14}\,cm^{-3}]}$\,GV/m. %
When adjusting the plasma density to satisfy $k_{pe}\sigma_z\cong1$, the estimate for the maximum longitudinal electric field can be expressed as $E_{WB}=\frac{m_ec^2}{e}\frac{1}{\sigma_z}$. %
With $\sigma_z$=12\,cm one obtains $E_{WB}\cong$\,27MV/m in a plasma with $n_e\cong8\times10^{10}\,cm^{-3}$~\cite{bib:caution1}. %

Producing a high-energy p$^+$ bunch with length shorter than $1\,mm$ to reach large wakefield amplitudes ($\sim$1\,GV/m as suggested in~\cite{bib:caldwellshort}) is very challenging. %
The usual magnetic compression methods would require a large additional energy spread ($\Delta E/E\cong1\%$) and a large magnetic chicane or drift space (km length)~\cite{bib:guoxing}. %

Reaching $E_{WB}>$1\,GV/m requires $n_e>10^{14}\,cm^{-3}$ and $\sigma_z<$\,5mm. %
In order to avoid transverse breaking up of the bunch because of the current filamentation instability (CFI)~\cite{bib:cfi}, the radius of the plasma, and thus its density, must be such that $k_{pe}\sigma_r\lesssim1$~\cite{bib:cfiexp}. %
In AWAKE the $p^+$/bunch can be focused to a rms radius $\sigma_r=200\,\mu m$, which requires a plasma density of n$_e\le7\times$10$^{14}$\,cm$^{-3}$ for $k_{pe}\sigma_r\le1$. %
In this case the bunch is many wakefields periods long and ineffective at driving wakefields. %

A bunch with $\sigma_z\gg\lambda_{pe}=2\pi/k_{pe}$ drives low amplitude wakefields with multiple $\lambda_{pe}$ periods, as was demonstrated in~\cite{bib:fang}. %
These wakefields have accelerating and decelerating longitudinal fields ($E_z$). %
When the bunch particles have a relativistic factor $\gamma>1$ their periodic energy gain and loss from the wakefields can lead to longitudinal bunching at the scale of $\lambda_{pe}$ only over large distances. %
In the AWAKE case the difference in travel distance between two particles of similar energy $\gamma_0$, one gaining and one losing energy from the wakefields and thus separated by $2\Delta\gamma\ll\gamma_0$ is given by $\Delta L\cong\frac{1}{\gamma_0^2}\frac{2\Delta\gamma}{\gamma}L\ll\lambda_{pe}$. %
Therefore longitudinal bunching from periodic energy gain/loss from the wakefields cannot be used to modulate the $p^+$ bunch charge or current. %

At the plasma entrance, the transverse fields ($E_{r}-v_bB_{\theta}\cong E_{r}-cB_{\theta}$) are focusing over most of the drive bunch length. %
In the transverse dimension, the particles have non-relativistic velocities whose rms value can be evaluated at a beam waist from the bunch normalized emittance $\epsilon_N$ as: $\sigma_{v_{\perp}}/c\cong\epsilon_N/(\gamma\sigma_r)$ and thus $\sigma_{v_{\perp}}/c\ll1$ when $\epsilon_N\ll1$. %
Therefore, even the weak transverse wakefields driven by the long bunch can lead to increase and decrease of the bunch density. %
In linear plasma wakefield theory, the wakefields amplitude is directly proportional to the (local) bunch density~\cite{bib:linPWFA}. %
Focused bunch regions thus drive larger wakefields and are in turn more focused, while other $p^+$ are defocused, which leads to the feedback mechanism for the self-modulation process~\cite{bib:kumar}. %
After saturation, the self-modulation has transformed the long bunch into a train of bunches, each shorter than $\lambda_{pe}$ and with period $\lambda_{pe}$. %
The formation of the bunch train is thus due purely to the transverse action of the wakefields. %
Assuming that the $k_{pe}\sigma_r\lesssim1$ condition is satisfied, the individual bunches satisfy $k_{pe}\sigma_z\le1$ and the train can thus resonantly and effectively drive the wakefields to large amplitude. %

Counting on the noise in the bunch and plasma as sources to initiate the modulation process would mean that the phase of the wakefields along the Gaussian $p^+$ bunch would be random and vary from event to event. %
This would make deterministic injection of electrons in the focusing and accelerating phase of the wakefields impossible. %
Therefore the self-modulation process must be seeded. %
Seeding also reduces the plasma length needed for the modulation process to saturate. %
Seeding methods include: a short laser or particle bunch preceding the long bunch and driving wakefields; a short particle bunch of charge opposite that of the drive bunch traveling within the drive bunch; a sharp "cut" in the front of the drive bunch distribution; a relativistic ionization front, created for example by a laser pulse traveling within the drive bunch; pre-modulation of the drive bunch. %
Seeding the wakefields at a level exceeding the noise level means that the seeded self-modulation (SSM) amplifies the initial wakefields rather than develops from random noise. %

\section{The AWAKE Experiment}

The AWAKE experiment at CERN~\cite{bib:edda} aims to study the driving of wakefields using $p^+$ bunches and to demonstrate the acceleration of electrons externally injected in these wakefields. %
Table~\ref{table:expparams} lists the general experimental parameters. %

\begin{table}[]
\centering
\caption{Beam, vapor, plasma and laser parameters of the AWAKE experiment.}
\begin{indented}
\label{table:expparams}
\item[]\begin{tabular}{@{}lllll}
\hline
\textbf{Parameter}                         & \textbf{Symbol}   & \textbf{Value}         & \textbf{Range}                & \textbf{Unit} \\ \hline
\textit{$p^+$ Bunch}     		   &                             &                               &-                                        &-                    \\ \hline
Energy                                          & W$_0$                  & 400                        &-                                         & GeV          \\ 
Relativistic Factor                         & $\gamma_0$         & 427                        &-                                         &-          \\ 
Population                                    & N                          & $3\times10^{11}$    &     $(1-3)\times10^{11}$            & $p^+$ /bunch \\ 
Length                                          & $\sigma_{z} $        & 12                        &-                                       & cm \\ 
Focused Size                               & $\sigma_{r} $        & 200                        &-                                       & $\mu$m \\ 
Normalized Emittance                  & $\epsilon_{N} $       & $3.5\times10^{-6}$  &-                                      & m-rad \\ 
$\beta$-function at Waist             & $\beta_{0}=\frac{\gamma_0\sigma_{r}^2}{\epsilon_{N}}$         & 5                              &-                                       & m \\ 
Relative Energy Spread               & $\Delta W_0/W_0$  & 0.03\%                   &-                                       &-  \\ \hline
\textit{Rubidium Vapor}               &  \textit{Rb}             &                                  &                                        &-                    \\ \hline
Density                                        & n$_{Rb}$            & $7\times10^{14}$   &   $(1-10)\times10^{14}$  & cm$^{-3}$          \\ 
Column Length                            & L$_{Rb}$            & 10                           &-                                       & m          \\ 
Column Radius                            & r$_{Rb}$             & 2                            &-                                       & cm          \\ \hline
\textit{Fiber/Ti:Sapphire Laser}                                  &  \textit{}&                               &                                        &                    \\ \hline
Central Wavelength                       & $\lambda_{0}$       & 780                       &-                                        & nm          \\ 
Bandwidth                                     & $\Delta\lambda_{0}$ & $\pm$5                   &-                                        & nm          \\ 
Pulse Length                                & $\tau_{0}$              & 120                       &-                                        & fs          \\ 
Max. Compressed Energy    & $E_{max}$              & 450                       &-                                        & mJ          \\ 
Focused Size                              & r$_{l}$                    & 1                            &-                                        & mm          \\ 
Rayleigh Length                           & Z$_{r}$                    & 5                       &-                                        & m          \\ \hline
\textit{Plasma}                              &                                &                               &-                                        &                    \\ \hline
Electron Density                           & n$_{e}$            & $7\times10^{14}$   &   $(1-10)\times10^{14}$  & cm$^{-3}$          \\ 
Electron Plasma Frequency                       & f$_{pe}$          & 237                      &90-284                          & GHz          \\ 
Electron Plasma Wavelength                      & $\lambda_{pe}$ & 1.3                  &3.3-1                                       & mm          \\ 
Length                                            & L$_{p}$            & 10                           &-                                       & m          \\ 
Radius                                             & r$_{p}$             & $>$1                            &-                                       & mm          \\ \hline
\end{tabular}
\end{indented}
\end{table}

In the following we describe the main elements of the AWAKE experiment, as well as the diagnostics that were implemented to characterize the modulated $p^+$ bunch exiting the plasma. %

\begin{figure}[ht]
\centering
\includegraphics[scale=0.6]{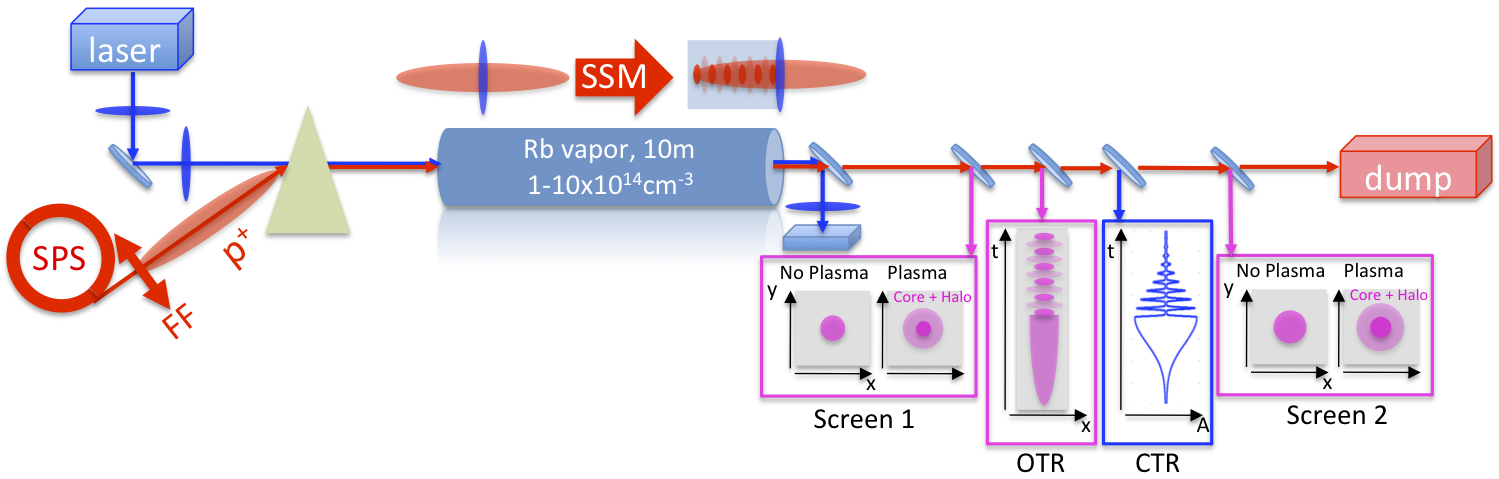}
\caption{Schematic of the AWAKE experiment.} 
\label{fig:AWAKEsetup2} 
\end{figure}

Figure~\ref{fig:AWAKEsetup2} shows a schematic of the AWAKE experiment. %
The $p^+$ bunch is extracted from the SPS with the parameters in Table~\ref{table:expparams}. %
It is put on the trajectory of the laser pulse by a bending dipole magnet that is part of a magnetic dogleg. %
The final focus system focuses the bunch near the entrance of the vapor source. %
The short laser pulse propagates within the $p^+$ bunch to form the plasma by ionization of the rubidium (Rb) vapor, thereby seeding the SSM (figure~\ref{fig:AWAKE3schemes} and see below). %
The SSM develops along the plasma. %
We measure the effect of the SSM on the $p^+$ bunch  downstream from the plasma. %

The development of the SSM and the growth of the wakefields lead to two observable effects on the $p^+$ bunch:
defocusing of some of the protons and modulation of the $p^+$ bunch density, reaching zero near the bunch axis in case of SSM saturation and large amplitude wakefields. %
Bunch diagnostics include: two screens to observe $p^+$ defocused along the SSM growth and to determine their angle and position of origin along the plasma; analysis of the spatio-temporal structure of the optical transition radiation emitted by the modulated $p^+$ bunch; spectral analysis of the coherent transition radiation emitted by the $p^+$ bunch train. %

\begin{figure}[ht]
\centering
\includegraphics[scale=0.6]{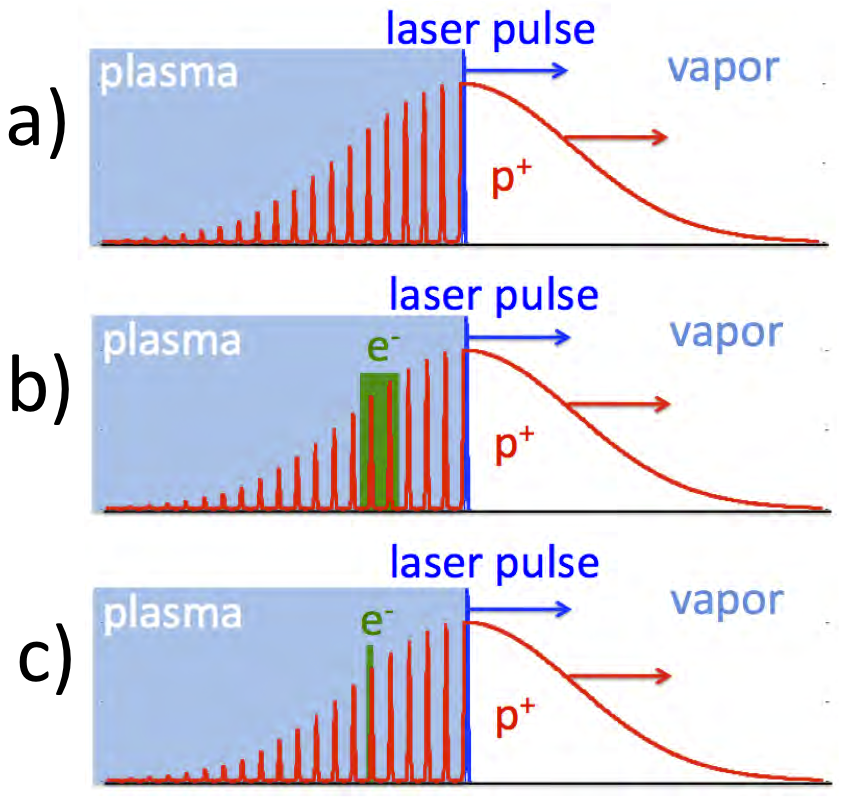}
\caption{The three schemes of the AWAKE experiment: a) $p^+$ bunch, Rb vapor and ionizing laser pulse to study the SSM; b) a long electron bunch ($\sigma_z=(1-2)\lambda_{pe}$) is injected and a fraction of the electrons is captured and accelerated; c) a short electron bunch ($\sigma_z\ll\lambda_{pe}$) is injected and accelerated with a narrow energy spread while preserving its initial emittance. The ratio $\sigma_z/\lambda_{pe}$ is much smaller than in AWAKE. } 
\label{fig:AWAKE3schemes} 
\end{figure}%

\subsection{Plasma Source}

The plasma source consists of a Rb vapor source and of a laser for ionization of the vapor. %

\subsubsection{Requirements}

The source must fulfill a number of requirements. %

The plasma electron density must be adjustable in the $(1-10)\times10^{14}$\,cm$^{-3}$ range. %

Numerical simulations show that at these densities the SSM takes a few meters to develop from its seed value. %
The plasma length must therefore be in the 10\,m range. %

The cold plasma skin depth is smaller than $\sim$530\,$\mu$m. %
The plasma radius must therefore be of the order of $\sim$1\,mm. %

The SSM can develop in plasmas with large density gradients~\cite{bib:schroederuniformity}. %
However, the injection and acceleration of an initially low energy electron bunch ($\cong$15\,MeV) are very sensitive to plasma density variations. %
Numerical simulations show that, with the seed at the peak of the $p^+$ bunch, the wakefields amplitude typically peaks approximately $\sigma_z$ after the ionization seeding location~\cite{bib:kumar}. %
That is, the location for injection of electrons along the bunch corresponds to $\sim100\lambda_{pe}$. %
Therefore, if the plasma density were to change by a relative amount $\delta n_e/n_{e0}$ with $\delta n_e\ll n_{e0}$, the plasma wavelength would change by a relative amount $\delta\lambda_{pe}/\lambda_{pe}=\frac{1}{2}\delta n_e/n_{e0}$. %
At the location of electron injection this change would be magnified by the number of plasma wavelengths $N_{\lambda_{pe}}$. %
Imposing that all along the plasma the electrons must remain in the accelerating and focusing phase of the wakefields, corresponding to a $\lambda_{pe}/4$ extent according to linear plasma wakefield theory, translates into a relative density uniformity estimate of $\delta n_e/n_{e0}<(1/4)(1/N_{\lambda_{pe}})$ all along the plasma. %
This leads to $\delta n_e/n_{e0}<0.25\%$ for $N_{\lambda_{pe}}=100$. %

At the same time, assuming the electrons are injected along the $p^+$ bunch trajectory, they must cross the plasma density ramps at the entrance and exit of the plasma. %
The electrons have a low energy at the plasma entrance and are sensitive to the transverse fields driven by the yet unmodulated $p^+$ bunch. %
Linear theory indicates that these fields are globally focusing for particles with the same sign as the drive bunch, i.e., defocusing for the electrons in this case. %
The length of the plasma density ramp must therefore be kept as short as possible, shorter than 10\,cm according to numerical simulations~\cite{bib:lotovramp}. %
The energy of the accelerated electrons exiting the plasma is much larger than their injection energy. %
They are thus much less sensitive to the defocusing effect of the wakefields in the exit ramp. %
However, the length of both ramps is similar. %

The plasma density must also be stable over time. %

\subsubsection{Rubidium Vapor Source}

\begin{figure}[ht]
\centering
\includegraphics[scale=0.6]{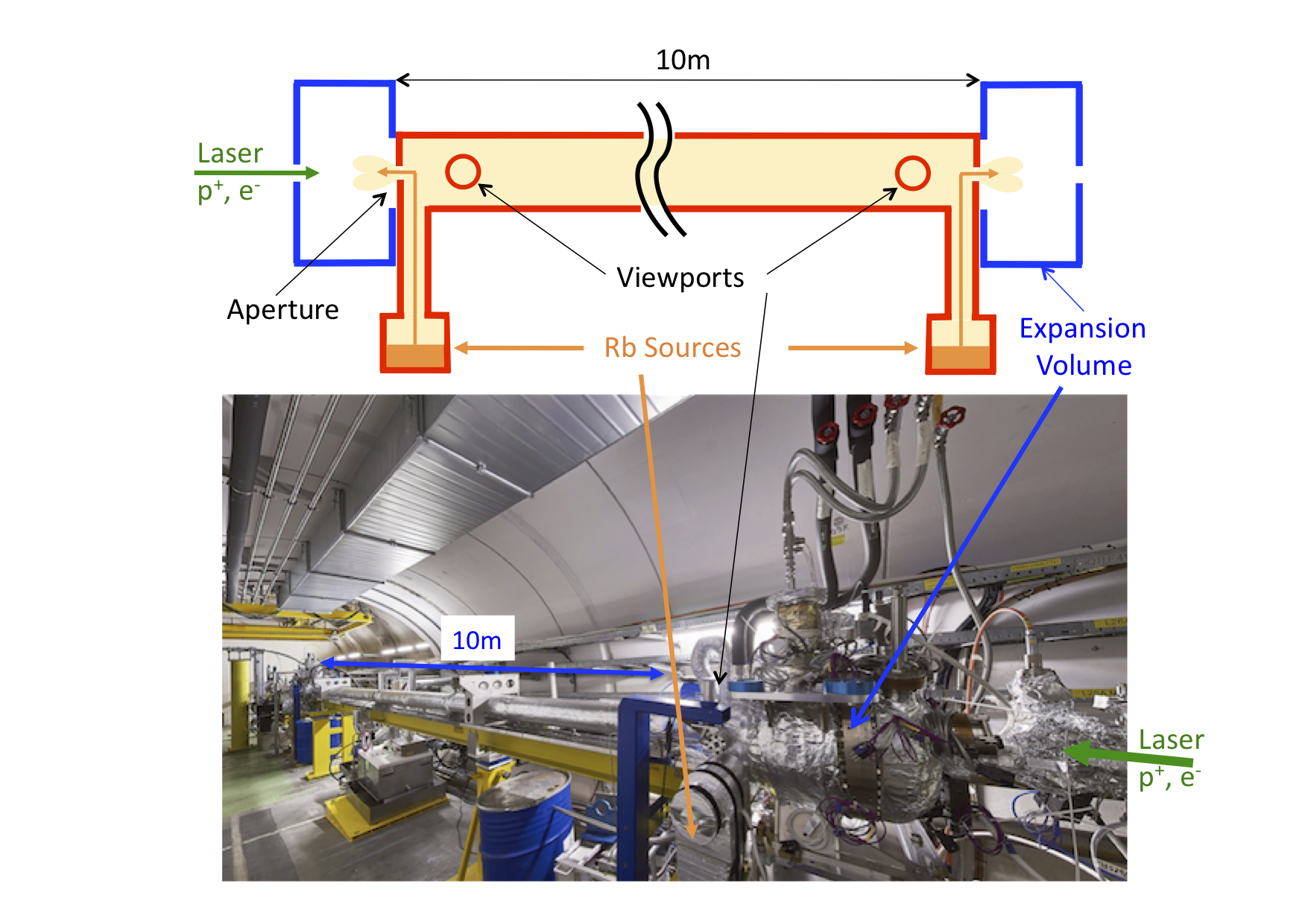}
\caption{Top: schematic of the Rb vapor source with the hot regions (Rb reservoirs, heat exchanger) in red-orange colors and the cold regions in blue color (expansion volume)~\cite{bib:vaporsource}. Bottom: picture of the actual source in the AWAKE experiment. } 
\label{fig:VaporSource} 
\end{figure}

The (neutral) vapor density of the Rb source~\cite{bib:vaporsource} developed for AWAKE (see figure~\ref{fig:VaporSource}) satisfies the requirements outlined in the previous paragraph~\cite{bib:erdemsource}. %
The source meets the density uniformity requirement by imposing with a heat exchanger a very uniform temperature ($\delta T/T<0.25\%$) and assuming no vapor flow along the 10\,m. %
The heat exchanger circulates an inert heat carrying fluid (Galden\textsuperscript{\textregistered} HT 270) in a 70\,mm tube surrounding the 40\,mm vacuum tube containing the Rb vapor. %
Measurements show a temperature uniformity better than 0.5\,K around 500\,K ($\delta T/T<0.1\%$), near the temperature necessary to reach the highest density expected in AWAKE. %
The AWAKE density range is obtained at temperature between 150 and 230$^\circ$C. %

Letting the Rb vapor expand in a vacuum volume with its walls maintained below the Rb condensation temperature of 39.48$^\circ$C leads to a density ramp scale length on the order of the diameter of the aperture through which the vapor expands. 
The diameter of the aperture at the entrance and exit of the vapor source is 1\,cm. %

Two reservoirs located near the ends of the source evaporate the Rb. %
After filling the source volume the vapor flows from each reservoir to its closest aperture, with a flow path in the source of only a few centimeters. %
White light interferometry near both ends of the source, where there is no flow, yields the Rb density with a relative accuracy of better than 0.2\%~\cite{bib:ozdensity}. %

The vapor source includes "shutters" that are inserted in the along the beam path, in the expansion volumes, and opened only for $\approx1\,s$ for each $p^+$ bunch event. %
These shutters collect the Rb vapor and decrease the amount of Rb condensing along the beam line vacuum elements upstream and downstream of the vapor source ends. %

The Rb vapor satisfies the density, uniformity and ramp length requirements. %
Assuming full ionization of the first Rb electron makes the plasma density also satisfy these requirements. %
Field ionization turns the vapor density into an equal plasma density: $n_e=n_{Rb}$. %

\subsubsection{Laser Ionization}

Rubidium has a low ionization potential for its first electron (4.177\,eV) and is thus easier to ionize than elements that are in gaseous phase at room temperature (e.g., Ar, He, etc.). %
The appearance intensity for ionization of a Rb atom is on the order of $1.7\times10^{12}$\,Wcm$^{-2}$~\cite{bib:augst}. %
The ionization potential for the second Rb electron is 27.28\,eV, which requires an approximately $\sim$455 times higher intensity to ionize. %
We expect no secondary ionization with the laser intensities available at AWAKE ($<1.2\times10^{14}\,Wcm^{-2}$). %
Rubidium has a large atomic mass (Z=85, 87 for the two isotopes present in $\sim$72\% and $\sim$28\% fractions in natural Rb), which contributes to minimizing the possible deleterious effects of plasma ions motion~\cite{bib:vieiraion}. %

A simple estimate including the energy necessary for ionizing a tube of Rb vapor 10\,m-long and 1\,mm in radius (volume $\sim$8\,cm$^{3}$) with a density of $10^{15}$\,cm$^{3}$ ($\sim8\times10^{15}$ atoms) and requiring that the laser pulse intensity is equal to the appearance intensity at the column end and  r=1\,mm to ionize the "last atoms" shows that an energy of $\sim$70\,mJ per laser pulse is necessary. %

The laser system consists of an erbium-doped fiber oscillator, frequency doubled to seed a Ti:Sapphire chirp-pulse amplification chain. %
The system produces pulses with characteristics given in Table~\ref{table:expparams}. %
They are $\sim$120\,fs-long, carry up to 450\,mJ and are focused to a peak intensity of $\sim1.2\times10^{14}$\,Wcm$^{-2}$ with a spot with radius 1\,mm and propagate with a Rayleigh length of $\sim$5m. %
The laser is operated at 10\,Hz to maintain optimal thermal characteristics and stability of the amplification chain. %
However, the last amplifier is pumped with a few microseconds delay with respect to the laser pulse, except when in synchronization with the $p^+$ bunch, i.e., once every $\sim$30\,s. %
The 10\,Hz repetition rate produces only $\sim$10\,mJ, which spares the optics and the vapor source components. %

The focused intensity exceeds the appearance intensity for the ionization of the first Rb electron. %
However, the laser spectrum overlaps the D2 optical transition of Rb at 780.2\,nm. %
Since this transition originates from the atom ground state, absorption and anomalous dispersion can strongly affect the laser pulse propagation, even in the low density vapor. %
At very low intensity, only a very small Rb vapor density-length (n$_{Rb}=6\times10^{14}$\,cm$^{-3}$ over 3.5\,cm) product is sufficient to stretch the short laser pulse by a factor of more than five and thereby reduce its intensity proportionally~\cite{bib:atefeh}. %

\begin{figure}[ht]
\centering
\includegraphics[scale=0.2]{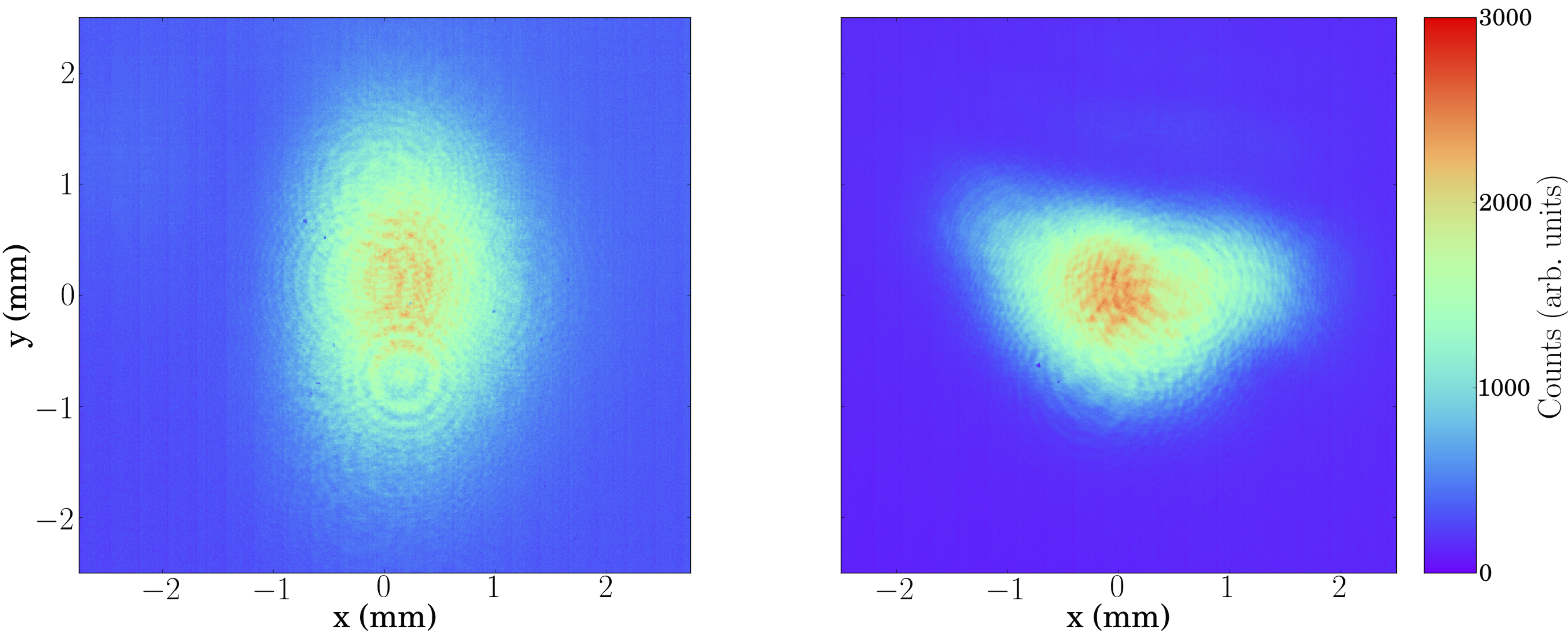}
\caption{Images of the laser pulse transverse profile obtained by imaging the pulse at the vapor source exit. a) Propagation with 200\,mJ in vacuum obtained without Rb vapor in the source, and b) with Rb vapor in the source with $n_{Rb}=7\times10^{14}\,cm^{-3}$ and 240\,mJ.} 
\label{fig:laserpropag}
\end{figure}

At high intensity, the front of the laser pulse may stretch in the vapor. %
However, when the pulse intensity exceeds the ionization threshold, the remainder of the pulse can propagate along the plasma because the plasma is a very weakly dispersive medium over the laser pulse bandwidth. 
Measurements of the transverse laser pulse profile (see figure~\ref{fig:laserpropag}) and auto-correlation indeed show that above approximately 100\,mJ,  the laser pulse propagates through the 10\,m vapor column. 
Figure~\ref{fig:laserpropag} shows that the laser pulse experiences some transverse distortion and this will be the topic of further experimental and simulation work. %
The propagation of the laser pulse is the indication of the depopulation of the Rb atom ground state. %
Ionization from the first excited state or from other states also populated by the laser pulse photons should occur at even lower intensity than from the ground state. %
Therefore, with sufficiently high laser pulse intensities, the resonant interaction between the laser pulse and the Rb atoms could actually enhance the ionization process. %

The timing between the $p^+$ bunch and the laser pulse can also be set such that ionization occurs ahead of the $p^+$ bunch. %
This would mimic the case of a preformed plasma. %
The plasma itself is produced at the time of the laser pulse. %
At each location and from the time of production, the plasma density can evolve through diffusion because of its finite transverse size, or through (mainly) three-body recombination. %
These processes have different evolution time scales. %
At the low plasma densities of AWAKE, the evolution is typically at the microsecond time scale (see measurements in Ref.~\cite{bib:wangthesis}). %

In this case one can answer a number of questions in the context of AWAKE. %
First, does the self-modulation instability occur even with no seed? %
Second, can we observe the competition predicted between the self-modulation instability and the hosing instability~\cite{bib:witthum,bib:vieirahosing}. 
Third, can the competition be suppressed (over the AWAKE plasma length) by seeding the self-modulation instability~\cite{bib:vieirahosing}. 

For example, the results obtained in Ref.~\cite{bib:vieirahosing} could be interpreted the following way. %
The part of the $p^+$ bunch behind the seed point experiences SSM in a first plasma, while the front of the bunch that propagates in a vapor remains unaffected. %
In a second, preformed and long plasma dedicated to acceleration of electrons, the front part of the bunch could be subject to a self-modulation instability (unseeded) or to the hosing instability, depending on the noise level for the two instabilities. %
This could have important implication for the use of a second, pre-formed plasma for acceleration over long distances. %
We note here that the natural divergence of the front of the bunch, not subject to plasma focusing, could mitigate the growth of instabilities in the front of the bunch. %

\subsection{Self-Modulation Seeding}

With the elements described above, the seeding of the SSM is possible. %
The part of the bunch ahead of the laser pulse does not interact with the neutral Rb vapor. %
The laser pulse that is itself shorter than the plasma period (120\,fs$\ll$4\,ps) and the ionization process that occurs over an even shorter time scale provide a good seed for the SSM~\cite{bib:vieraseed}. %
The sudden onset of the full plasma density in the middle of the $p^+$ bunch is equivalent to the sudden onset of the $p^+$ bunch in a preformed plasma. %
This equivalence is used to evaluate the seed wakefields amplitude (see below) and in numerical simulations so that neither the full bunch, nor the ionization process need to be simulated by using a "cut" $p^+$ bunch. %

Assuming the laser pulse propagates along the 10\,m in a plasma with density 10$^{15}$\,cm$^{-3}$, its velocity is given by its group velocity $v_g\cong\left(1-\frac{1}{2}\frac{\omega_{pe}^2}{\omega_l^2}\right)c$ (for $\omega_{pe}^2/\omega_l^2\ll1$, $\cong6\times10^{-7}$ here). %
The $p^+$ bunch propagates with a velocity given by its relativistic factor: $v_b\cong\left(1-\frac{1}{2\gamma^2}\right)c$ (for $\frac{1}{\gamma^2}\ll1$, $\cong5\times10^{-6}$ here). %
The dephasing distance between the two and over the plasma length is thus: $\Delta L\cong\left(\frac{v_g}{v_b}-1\right)L\cong\left(\frac{1}{2\gamma^2}-\frac{\omega_{pe}^2}{\omega_l^2}\right)L\cong50\mu m\ll\lambda_{pe}$. %
Therefore, de-phasing between the laser pulse and the $p^+$ bunch driving the wakefields is not an issue for the seeding of the SSM. %
In addition, evolution of the laser pulse that does not decrease its intensity below the ionization intensity does not affect the seeding either since the laser pulse is so much shorter than the typical period of the wakefields. %

We evaluate the initial wakefields driven at the plasma entrance from PWFA linear theory~\cite{bib:linPWFA}. %
With the AWAKE parameters ($\sigma_z\ll\lambda_{pe}$) the bunch density can be considered constant over a few plasma periods and the peak initial wakefield amplitude can be written as:%
\begin{equation}\label{eq:wz}
W_z=\frac{e}{\epsilon_0k_{pe}}n_{b0}\cdot R(r),
\end{equation}
\begin{equation}\label{eq:wz}
W_{\perp}=2\frac{e}{\epsilon_0k_{pe}^2}n_{b0}\cdot dR(r)/dr,
\end{equation}
where $n_{b0}$ is the bunch density at the seeding point. %
The terms $R(r)$ and $dR/dr$ are geometric factors, which for $k_{pe}\sigma_r=1$ take the values $R(r=0)\cong0.46$ and $dR(r=\sigma_r)/dr=-958\,m^{-1}$. %
This leads to initial values of 6.2 and 5.6\,MV/m for $W_z$ and $W_{\perp}$, respectively. %
Those amplitudes are much larger that the tens of keV values expected from noise~\cite{bib:lotovseed} and demonstrates the effectiveness of the seeding. %
Numerical simulations show that the wakefields grow from the seeded values to hundreds of MV/m values in $\sim$4\,m~\cite{bib:edda}. %
Therefore this seeding of the instability makes it possible to amplify the wakefields with well defined initial phase and amplitude. %

Numerical simulations also show that, initially, the phase velocity of the wakefields is lower than that of the drive bunch and of the laser pulse~\cite{bib:kumar}. %
The slow down is the result of the evolution of the $p^+$ bunch and the growth of the wakefields. %
This means that the phase of the wakefields at the point where the electrons would be externally injected for acceleration, that is, the phase of the wakefields relative to the laser pulse changes so much that the electrons find themselves in the wrong phase of the wakefields and probably move out of the wakefields and are lost through defocusing. %

\begin{figure}[ht]
\centering
\includegraphics[scale=0.4]{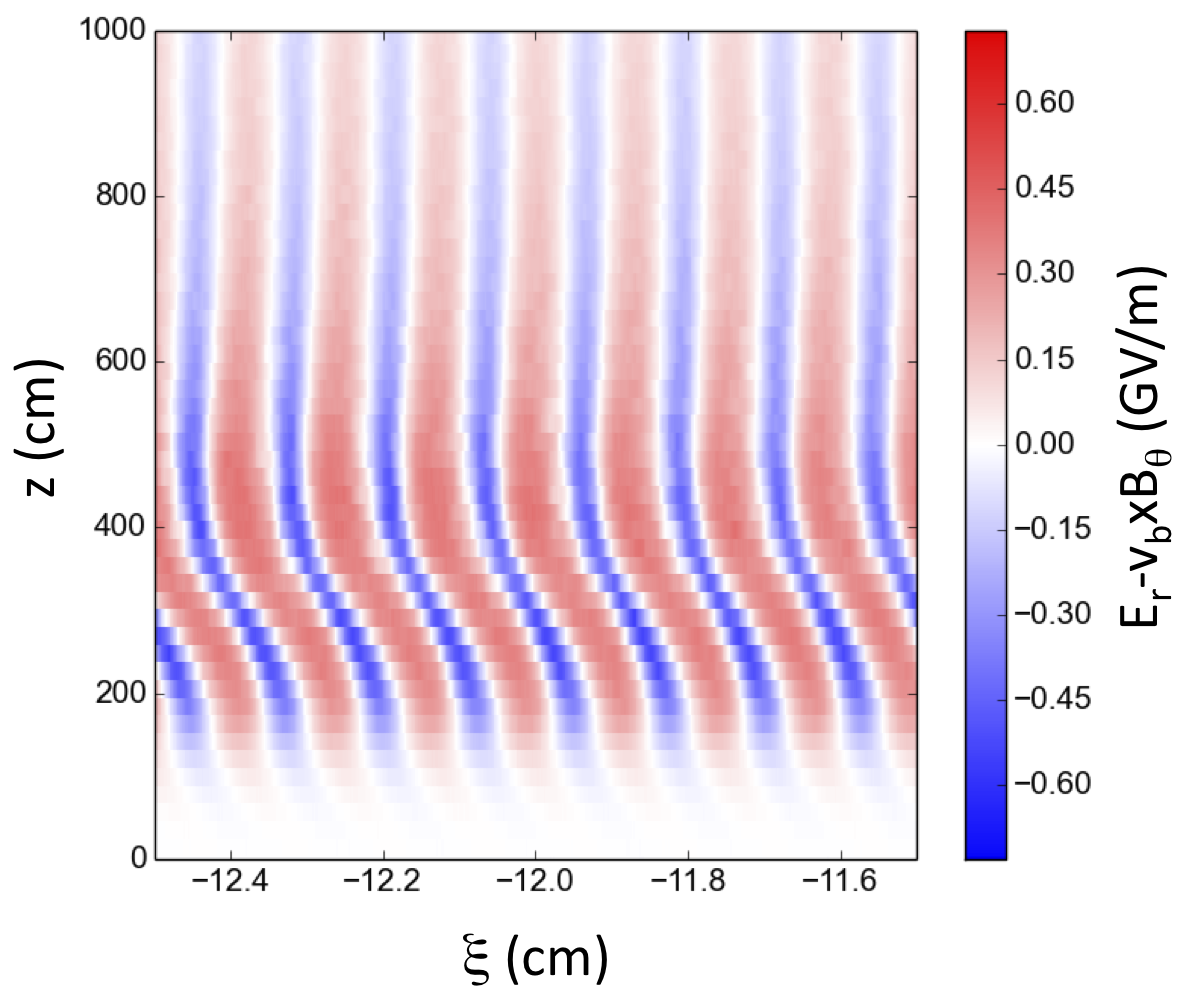}
\caption{Plot of the evolution of the transverse wakefields in phase and amplitude, focusing for protons in blue, along the bunch $\xi$ and along the plasma z, around the point leading to the largest energy for electrons $\xi\cong$12\,cm. Parameters are those of AWAKE with $n_e=7\times10^{14}\,cm^{-3}$.} 
\label{fig:WperpfigNicolas2} 
\end{figure}

Figure~\ref{fig:WperpfigNicolas2}  (from Ref.~\cite{bib:savard}) shows the typical evolution of the wakefields along the plasma near the injection point, a distance $\xi\cong\sigma_z$=12\,cm behind the laser pulse. %
The figure is obtained from the moving window in 2D cylindrically symmetric PIC simulations using the code OSIRIS~\cite{bib:osiris}. %
It confirms the expected evolution of the wakefields along the plasma. 
The transverse wakefields amplitude ($E_r-v_b\times B_{\theta}\cong E_r-c\times B_{\theta}$), i.e., the transverse force per unit charge is evaluated at the initial radius of the bunch, 200\,$\mu$m. %
The wakefields grow from initially focusing all along the bunch, to alternatively focusing/defocusing. %
During growth, the wakefields shift backwards in the moving window (itself moving at c) until z=4-5\,m. %
Their phase remains essentially constant after that, i.e., after saturation of the SSM process. %
The longitudinal wakefields amplitude E$_z$ follow the same trend (not shown). %
Therefore, electrons injected after z=4-5\,m remain in phase with the wakefields and can be accelerated over long distances and to large energies. %
More importantly, simulations show that the relative phase of the wakefields with respect to the moving window and the seed of the wakefields after z=4-5\,m varies by less than $\lambda_{pe}/10$ with $n_e=7\times10^{14}\,cm^{-3}$ and under variations of the bunch parameters N,$\sigma_r$,$\sigma_z$ by $\pm$5\%~\cite{bib:savard}. %

These results show that the strong seeding by the ionization front fixes the phase of the wakefields reached after saturation with variations that allow for external injection of electrons with the accelerating and focusing phase of the wakefields. %

These results are also the seed for a study for AWAKE Run 2, whose goal is the acceleration of an externally injected electron bunch to a high energy (multi-GeV) with a narrow final energy spread and with preservation of the incoming bunch emittance (at the mm-mrad level). %
For these experiments the plasma could be split between a self-modulation and an acceleration section. %
The self-modulated plasma should be at least 4\,m long. %
The electrons can then be injected from between the two plasmas and into the acceleration plasma where the evolution of the wakefields is expected to be less than during the development of the SSM. %
Since no seeding is required in the second plasma, the second source could be of a different kind to the first one. %
However, possible growth of the self-modulation instability in the front of the $p^+$ bunch and from noise, as well as of other competing instabilities should be studied carefully before adopting this scheme. %

The current set up also allows for the study of the evolution of the $p^+$ bunch in a preformed plasma, that is without seed for either the self-modulation or the hosing instability. %
This can be simply realized by placing the laser pulse an arbitrary distance ahead of the $p^+$ bunch. %

\section{SSM Diagnostics}

The diagnostics are currently geared towards the effects of the SSM on the $p^+$ bunch, as outlined earlier: protons defocusing and modulation of the bunch density. %

\subsection{Protons Defocusing Measurements}

The proton bunch is focused near the entrance of the plasma. %
Without plasma and with $\beta_0=5\,m$ the beam transverse rms size is 520\,$\mu$m at a screen located 2\,m downstream and 830\,$\mu$m at a screen located 10\,m downstream from the plasma exit. %
Without plasma, the bunch transverse profile is Gaussian at these two locations. %
Protons focused by and driving the wakefields exit at the end of the plasma and travel ballistically after that. %
Defocused $p^+$ may leave the wakefields before the end of the plasma and travel ballistically after that and form a halo around the focused core. %
A focused core and a halo on the transverse bunch images are evidence of the development of the SSM. %
With two screens the ballistic trajectory of the defocused protons can be reconstructed and the location along the plasma where they exited the wakefields can in principle be determined~\cite{bib:turner}. %
This measurement could yield important information about the saturation point of the SSM along the plasma as a function of the bunch and plasma parameters (n$_e$, N, $\sigma_z$, etc.), a fundamental property of the SSM. %

\subsection{Proton Bunch Modulation Measurements}

In its initial stage of growth, the SSM leads to modulation of the bunch radius. %
However, as the growth continue the SSM reaches a nonlinear stage in which the wakefields reach a significant fraction of the cold wave-breaking field. %
The plasma density perturbation sustaining these wakefields is also non-linear and so is the corresponding $p^+$ bunch density modulation. %

When the $p^+$ bunch enters a metal foil placed downstream from the plasma exit it emits (backward) transition radiation (TR). %
This TR has a very broad spectrum, in theory from zero frequency to the metal equivalent electron plasma frequency, even in the x-ray range near atomic transitions. %
For a nanosecond bunch with time structure at the few picosecond scale and a radius of 200\,$\mu$m, the TR emitted in the visible range by the bunch ensemble is incoherent and is called optical transition radiation (OTR). %
Since the TR emission is prompt, the OTR light has the same time and spatial structure as the $p^+$ and can thus be imaged and time resolved. %

The radiation corresponding to the charge density modulation period resulting from the SSM is emitted coherently and is referred to as coherent transition radiation (CTR). %
With the few picosecond period time scale it corresponds to microwaves in the 90-284\,GHz frequency range. %
Its frequency can be analyzed using standard microwave techniques such as high-pass filtering from waveguides cut-off and heterodyne mixing. %

\subsubsection{OTR and Streak Camera Measurements}

\begin{figure}[ht]
\centering
\includegraphics[scale=0.6]{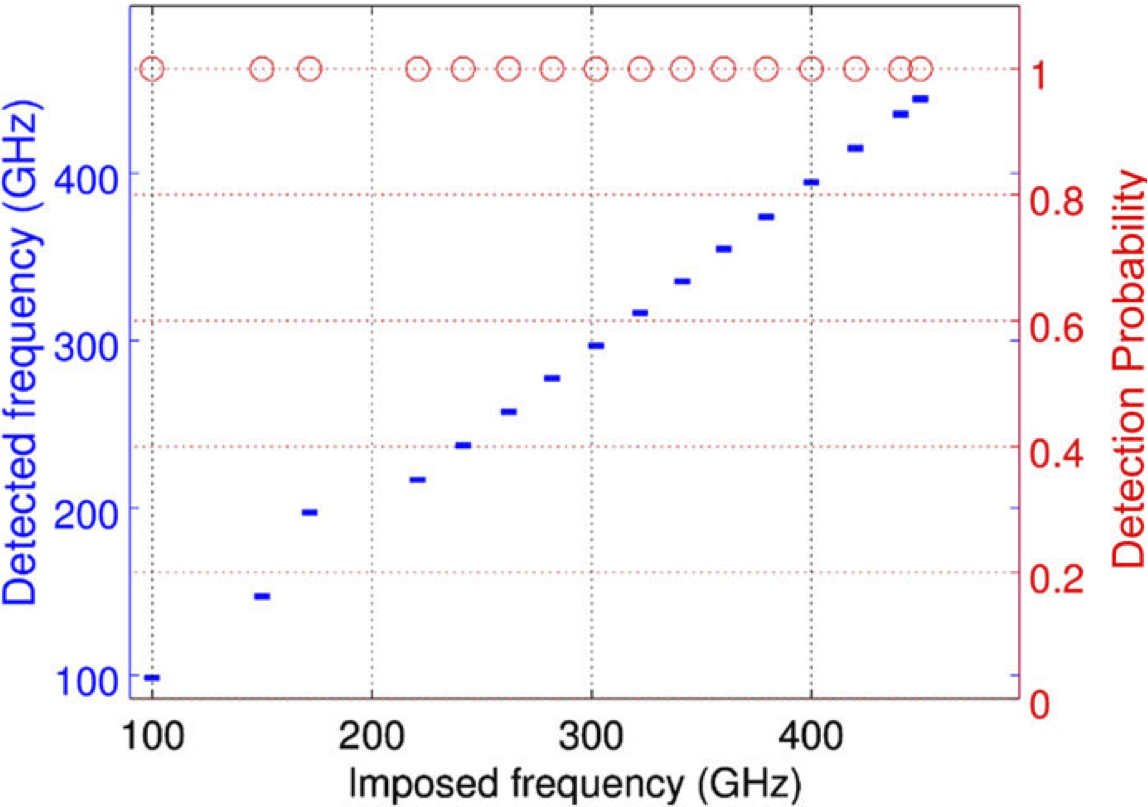}
\caption{Detection probability (open circles) and mean detected frequency (dashes). A one percent false detection threshold allows the detection of frequencies up to 450\,GHz. From Ref.~\cite{bib:rieger}.} 
\label{fig:steakresp} 
\end{figure}

\begin{figure}[ht]
\centering
\includegraphics[scale=0.3]{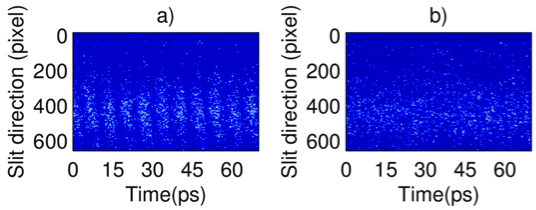}
\caption{Images obtained during the streak camera tests with modulation frequencies of a) 150 and b) 450\,GHz. From Ref.~\cite{bib:rieger}.} 
\label{fig:teststreakimages} 
\end{figure}

\begin{figure}[ht]
\centering
\includegraphics[scale=0.9]{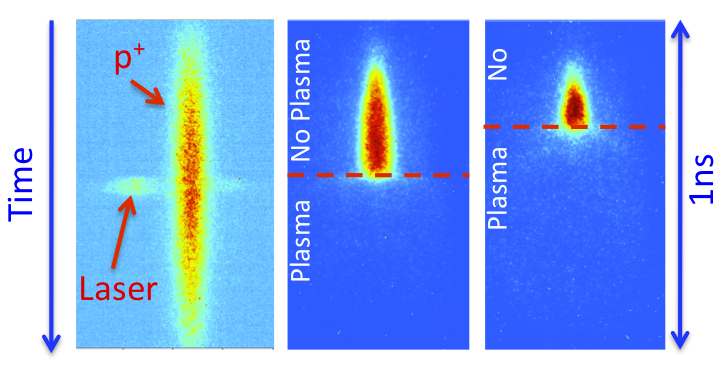}
\caption{Streak camera images at the nanosecond time scale. Left image: $p^+$ bunch and laser pulse at low power (no plasma). Middle image: $p^+$ bunch with laser pulse at the same time as in the right image, but with high energy and blocked  (see figure~\ref{fig:AWAKEsetup2}) not to damage the streak camera (and thus not visible), creating a plasma for the part of the bunch behind of the laser pulse. The effect of the plasma on the bunch is clearly visible, the laser time or position is indicated by the red dashed line. Right: image similar to the middle image, but with the laser pulse earlier in the bunch demonstrating the ability to trigger the plasma at any time along the $p^+$ bunch.} 
\label{fig:StreakSeeding} 
\end{figure}

We demonstrated that a streak camera with picosecond impulse response can resolve the modulation frequency of a ns-long light pulse, i.e., with a time structure similar to that expected from the $p^+$ bunch OTR~\cite{bib:rieger}. %
The modulation of the input light was obtained by beating in an optical fiber two CW diode laser beams with precisely tunable frequencies. %
An optical fiber modulator gated the beat signal to give it a nanosecond length, similar to that expected from the OTR signal. %
Figure~\ref{fig:steakresp} shows that the probability for measuring the modulation frequency was very high, all the way to up to at least 450\,GHz. %
Figure~\ref{fig:teststreakimages} shows two streak images obtained during these measurements. %
Figure~\ref{fig:teststreakimages} a) shows that for low frequencies ($<$150\,GHz), the modulation is directly visible on the image. %
Fourier analysis confirms the frequency of the modulation visible on the image and was used for figure~\ref{fig:steakresp}. %
For higher frequencies ($<$300\,GHz) the modulation is not visible (figure~\ref{fig:teststreakimages}(b)), but Fourier analysis yields the expected modulation frequency value, as figure~\ref{fig:steakresp} shows. %
For even higher frequencies, the average of multiple Fourier power spectra yields the expected frequency. %
We note here that the test images were acquired at kHz repetition rate. %
In the initial AWAKE experiment, the $p^+$ bunch is delivered to the plasma every $\sim$30\,s. %
However, since the vapor and plasma source are thermal systems with large mass, the plasma density and thus the modulation frequency are expected to remain very constant over the time scale of tens of $p^+$ bunch events. %
Averaging of the Fourier signals is thus appropriate. %
Figure~\ref{fig:StreakSeeding} shows streak images obtained with the SPS $p^+$ bunch and the ionizing laser pulse. %
The first image shows that the $p^+$ bunch and the laser pulse can be synchronized and the laser pulse put in the middle of the $p^+$ bunch for the seeding of the SSM. %
Synchronization uses a 6\,GHz master clock locking in phase both the laser oscillator at $\sim$88\,MHz and the SPS RF system frequency at $\sim$200\,MHz~\cite{bib:heiko}, and synchronizing the SPS extraction with the 10\,Hz repetition rate of the laser amplification chain. %
The rms length of the $p^+$ bunch is expected to be $\sim$400\,ps and the image shows $\sim\pm2\sigma_z$ of the bunch. 
The laser pulse appears longer than it is ($\sim$120\,fs) because of the limited time resolution of the streak camera at this (slow) time scale. %
The second image is similar to the first image, except that the laser pulse now carries enough energy to ionize the Rb vapor and is thus blocked by the laser beam dumps (see figure~\ref{fig:AWAKEsetup2}) and not visible. %
However, it is clear that the $p^+$ bunch is strongly affected at times after the laser pulse, when it propagates in the plasma. %
The last figure shows that the plasma creation and the SSM can be triggered at an earlier time. %
Detailed results on the effect of the plasma on the $p^+$ bunch and the seeding of the SSM are in preparation. %

The $p^+$ bunch has an arrival time jitter of $\pm$15\,ps with respect to the laser pulse. %
This jitter is caused by synchrotron oscillations of the $p^+$ bunch with respect to the SPS accelerating RF wave. %
However, the growth of the SSM is slow when compared to that jitter and this is therefore not expected to cause variation in the wakefields phase. %

\subsubsection{CTR Measurements}

The transition between coherent and incoherent emission occurs at frequencies such that $\omega\tau\cong1$, where $\tau=\sigma_z/c$ in the case of the unmodulated bunch and $\tau=\lambda_{pe}/c$ in the case of a self-modulated bunch. %
Therefore the unmodulated bunch emits CTR at frequencies lower than $\sim$40\,GHz. %
Modulation frequencies are between 90 and 284\,GHz for the AWAKE plasma density range. %
The bunch train emits CTR with a spectrum centered around the modulation frequency. %
The low frequency component corresponding to the bunch charge envelope is filtered out using a section of waveguide cut-off for frequencies below 60\,GHz. %
The CTR is emitted in a cone of radially polarized radiation. %

Detailed calculations with proton bunch distributions from beam plasma simulations have been performed to determine the expected CTR parameters~\cite{bib:martyanov}. %
The angle of the cone depends on the modulation frequency, i.e., the plasma density, and decreases with increasing plasma density. %
The peak CTR intensity is in the 10-100\,Wcm$^{-2}$ at the location of the off-axis parabola used to collimate the CTR cone. %

The CTR frequency (f$_{CTR}$) can be analyzed simply with Schottky diodes placed after cut-off waveguides. %
Suitable standard microwave band cut-off frequencies f$_{co, WR-08-05-03}$ we use are 90 (WR-08), 140 (WR-05) and 220\,GHz (WR-03). %
A system of three diodes behind a centimeters-long waveguide and horn in each band is placed in the CTR emission cone with the proper orientation, that is with the small side of the waveguide horn along the CTR cone radius. %
The amplitude of the signal from these diodes directly yields information about the CTR frequency content, for each event. %
For example, when operation at low plasma density (n$_e<2.5\times10^{14}$cm$^{-3}$) such that  f$_{CTR}$=f$_{mod}<$f$_{co, WR-05}$ a signal on the WR-08 diode could be the sum of signals at all harmonics of f$_{mod}$ (weighted by the Schottky diode response at the various harmonic frequencies). %
Similarly, a signal on the WR-05 diode with (2.5$\times10^{14}<$n$_e<6\times10^{14}$\,cm$^{-3}$) could be the sum of signals from the second and above harmonics of f$_{mod}$. %
And a signal on the WR-03 diode could be the sum of signals from the third and above harmonics of f$_{mod}$. %
Considering that the signal  strength of each harmonic should decrease with increasing harmonic number, harmonic signal ratios can be obtained from each event and as a function of the experimental parameters (n$_e$, N, ...). %
With absolute calibration of the detection system, an estimate of the bunch radial modulation depth, and, in the case of strong modulation, an estimate of the micro-bunch length with respect to $\lambda_{pe}$ could be obtained. %

For higher plasma densities the system only yields information about the first and second harmonics (f$_{co, WR-05}<f_{mod}<f_{co, WR-03}$) or even only about the first (f$_{co, WR-03}<f_{mod}$) harmonic. %

We obtain precise measurements of the CTR and modulation frequency from heterodyne measurement systems in the WR-08 and WR-03 waveguide bands. %
The systems mix the CTR signal of unknown (but guessed from $n_{Rb}$ assuming full laser ionization) frequency $f_{CTR}$ (usually referred to as $f_{RF}$), with a signal of known frequency $f_{LO}$. %
The systems generate the frequency difference, called the intermediate frequency: $f_{IF}=\left|f_{RF}-f_{LO}\right|$. %
The value of $f_{LO}$ is chosen to bring $f_{IF}$ into the bandwidth of a fast oscilloscope: f$_{IF}\le$20\,GHz. %
We then use Fourier or wavelet analysis to determine the frequency content of the signal. %
Since the emission lasts for a time on the order of the half $p^+$ bunch length (400-800\,ps), the frequency difference must be kept large enough to obtain a few periods of $f_{IF}$  in the signal. %
Measurements with two slightly different values of $f_{LO}$ lift the ambiguity in frequency introduced by the absolute value in the expression for $f_{IF}$. %

We operate the heterodyne systems simultaneously by splitting the CTR signal and also obtain harmonic content information, in this case avoiding contributions of multiple harmonics to a single signal since the signals are frequency resolved. %
We also developed a heterodyne system for which the local oscillator signal is generated directly on the mixing diode by two CW infrared laser pulse tuned to beat at $f_{LO}$~\cite{bib:mishaheterodyne}, the same laser system used for the streak camera tests~\cite{bib:rieger}. %

\section{A look at the future}

Current experiments focus on the study of the SSM physics and span till the end of 2017. %
These require only the $p^+$ bunch, the Rb vapor source, the ionizing laser pulse and the diagnostics described here (see figure~\ref{fig:AWAKE3schemes}(a)). %

An electron beam injector capable of producing 10-20\,MeV electrons is in its installation phase. %
The electron bunch length is $\sim$10\,ps, long when compared to the period of the wakefields so that precise timing between the bunch and the wakefields is not critical for capture and acceleration. %
However, in this case only a fraction of the electrons are captured. %
The electron bunch is injected at the peak amplitude of the wakefields, $\sim\sigma_z$ behind the ionization laser pulse, as suggested by figure~\ref{fig:AWAKE3schemes}(b). %

The RF-gun is driven by a laser pulse derived from the oscillator of the ionizing laser system. %
Therefore, the jitter between the electron bunch and laser arrival time at the plasma entrance is expected to be $\sim$100\,fs. %
A streak camera with sub-ps resolution capturing light from the bunch and the laser pulse at a screen before the plasma will be used to measure the laser to electron bunch time jitter. %
The streak camera used to capture the beam modulation will be fitted with an optical fiber and a delay line to match the actual ionizing laser pulse propagation time of flight, so that a reference laser pulse can be made visible on each streak image as a relative time reference. %
In particular it will be free of the $\sim$20\,ps inherent jitter of the streak camera with respect to its trigger. %
This reference will be used to measure the stability of the phase of the wakefields with respect to the ionizing laser. %
This is an extremely important measurement that should demonstrate that the final phase of the wakefields with respect to the ionizing laser is smaller than a plasma period, as suggested by simulations~\cite{bib:savard}, and thus the effectiveness of the wakefields seeding. %
The position of the injected electrons with respect to the wakefields will thus be determined. %
These experiments will be performed first at low plasma densities for which the modulation of the $p^+$ bunch density, akin to that of the plasma electron density, can be directly observed on the streak camera images together with the accelerated electrons. %
At higher densities, only the modulation frequency can be determined for the Fourier transform of the bunch density and the phase information is lost. %

These external injection experiments will sample the wakefields driven by the $p^+$ bunch. %
A narrow final energy spread may be obtained because of the electron capture process~\cite{bib:edda}. %

In order to reach a high capture efficiency (100\%), a narrow final energy spread ($\sim1\%$) and preserve the emittance of the incoming bunch ($\sim$2\,$\mu$m), the injected electron bunch must be short ($\sim100\,fs<\lambda_{pe}/c$) and carry a large enough charge to load the wakefields. %
This is suggested in figure~\ref{fig:AWAKE3schemes}(c). %
In particular, with the parameters of Table~\ref{table:expparams} the plasma density perturbation does not reach the initial density. %
That means that plasma electrons remain in the wakefields structure, as in the linear regime. %
The transverse focusing force acting on the accelerated electrons is thus not linear with radius. %
The emittance of the witness bunch is not preserved. %

However, beam loading can at the same time flatten the wakefields for a narrow final energy spread and evacuate the remaining electrons. %
In this situation, the bulk of the electron bunch propagates in an ion column void of plasma electrons and its emittance can be preserved, at least against geometric aberrations~\cite{bib:veronica}. %

\section{Summary}
AWAKE is a proton-driven plasma wakefield acceleration experiment at CERN. %
Initial experiments focus on the seeded self-modulation of the 400\,GeV proton bunch in a 10\,m-long rubidium plasma with electron density adjustable in the 1 to 10$\times10^{14}$\,cm$^{-3}$ range. %
We show that the experimental setup briefly described here is ready for systematic study of the seeded self-modulation. %
We show that the short laser pulse used for ionization of the rubidium vapor propagates all the way along the column, suggesting full ionization of the vapor. %
We show that ionization occurs along the proton bunch, at the laser time and that the plasma that follows affects the proton bunch. %
We obtain evidence of seeded self-modulation occurrence from its effect on the proton bunch: defocusing of protons, charge modulation along the bunch and emission of coherent transition radiation at the modulation frequency. %
Preliminary results show clear evidence of seeded self-modulation, in accordance with expectations. %
We are analyzing the results that will be published elsewhere. %
Future experiments will aim at accelerating a low energy, externally injected electron bunch. %
We are also developing plans for acceleration of an electron bunch resulting in low relative energy spread and preservation of incoming emittance. %
These experiments will most likely use two plasma sources, a short one ($<$10\,m) for the seeded self-modulation to occur and one for acceleration only. %
The second source can in principle be very long (tens of meters) and we are exploring the possibility of using discharge or helicon plasma sources for these experiments. %
We are also exploring the particle physics cases that could be better addressed with this plasma-based accelerator than with a conventional accelerator~\cite{bib:mathew}. %

\section{Acknowledgements}
TRIUMF contribution was supported by NSERC and CNRC. %
This work was supported in parts by: EU FP7 EuCARD-2, Grant Agreement 312453 (WP13, ANAC2); and EPSRC and STFC, United Kingdom.  %
The support of DESY, Hamburg and the Alexander von Humboldt Stiftung is gratefully acknowledged. %
Contribution of the Budker INP team was supported by RSCF grant No 14-50-00080. %

\pagebreak


\begin{thebibliography}{99}
\bibitem{bib:caution1}Caution has to be taken with this reasoning. Additional considerations come from the fact that at the lower density the bunch density is larger than the plasma density and the linear theory argument is not strictly valid. In addition the finite transverse size of the bunch should also enter the evaluation of the wakefields amplitude. However, the argument remains globally valid, the expected wakefields are indeed much higher in the high plasma density case.
\end{thebibliography}
\end{document}